%% file: main.tex
\documentclass[sigconf]{acmart}

\usepackage{booktabs} 
\usepackage{multirow}
\usepackage{changes}
\usepackage{adjustbox}
\usepackage{tablefootnote}
\usepackage[utf8]{inputenc} 
\usepackage{graphicx}
\usepackage{hyperref}
\usepackage{subfigure}
\usepackage{pifont}
\usepackage{footnote}

\newcommand{\cmark}{\ding{51}}%
\newcommand{\xmark}{\ding{55}}%

\setcopyright{none}
\renewcommand\footnotetextcopyrightpermission[1]{}
\acmConference[in Review for KCap'19]{in Review for KCap '19: International Conference on Knowledge Capture}{Nov 19--22, 2019}{Marina del Rey, CA}

\begin{document}

\title[Open Research Knowledge Graph]{Open Research Knowledge Graph: Next Generation Infrastructure for Semantic Scholarly Knowledge}

\author{Mohamad Yaser Jaradeh}
\orcid{0000-0001-8777-2780}
\affiliation{%
  \institution{L3S Research Center, Leibniz University of Hannover}
}
\email{jaradeh@l3s.de}

\author{Allard Oelen}
\orcid{0000-0001-9924-9153}
\affiliation{%
  \institution{L3S Research Center, Leibniz University of Hannover}
}
\email{oelen@l3s.de}

\author{Kheir Eddine Farfar}
\affiliation{%
  \institution{TIB Leibniz Information Centre for Science and Technology}
}
\email{kheir.farfar@tib.eu}

\author{Manuel Prinz}
\orcid{0000-0003-2151-4556}
\affiliation{%
  \institution{TIB Leibniz Information Centre for Science and Technology}
}
\email{manuel.prinz@tib.eu}

\author{Jennifer D'Souza}
\orcid{0000-0002-6616-9509}
\affiliation{%
  \institution{TIB Leibniz Information Centre for Science and Technology}
}
\email{jennifer.dsouza@tib.eu}

\author{G\'abor Kismih\'ok}
\orcid{0000-0003-3758-5455}
\affiliation{%
  \institution{TIB Leibniz Information Centre for Science and Technology}
}
\email{gabor.kismihok@tib.eu}

\author{Markus Stocker}
\orcid{0000-0001-5492-3212}
\affiliation{%
  \institution{TIB Leibniz Information Centre for Science and Technology}
}
\email{markus.stocker@tib.eu}

\author{S\"oren Auer}
\orcid{0000-0002-0698-2864}
\affiliation{%
  \institution{TIB Leibniz Information Centre for Science and Technology, L3S Research Center, Leibniz University of Hannover}
}
\email{auer@tib.eu}

\renewcommand{\shortauthors}{M.Y. Jaradeh et al.}

\begin{abstract}
  Despite improved digital access to scholarly knowledge in recent decades, scholarly communication remains exclusively document-based. In this form, scholarly knowledge is hard to process automatically. In this paper, we present the first steps towards a knowledge graph based infrastructure that acquires scholarly knowledge in machine actionable form thus enabling new possibilities for scholarly knowledge curation, publication and processing. The primary contribution is to present, evaluate and discuss multi-modal scholarly knowledge acquisition, combining crowdsourced and automated techniques. We present the results of the first user evaluation of the infrastructure with the participants of a recent international conference. Results suggest that users were intrigued by the novelty of the proposed infrastructure and by the possibilities for innovative scholarly knowledge processing it could enable.
\end{abstract}

\keywords{Information Science, Knowledge Graph, Knowledge Capture, Research Infrastructure, Scholarly Communication}

\maketitle

\input{body}

\begin{acks}
This work was co-funded by the European Research Council for the project ScienceGRAPH (Grant agreement ID: 819536) and the TIB Leibniz Information Centre for Science and Technology. The authors would like to thank the participants of the ORKG workshop series, for their contributions to ORKG developments. We also like to thank our colleagues Kemele M. Endris who coined the DILS experiment, Viktor Kovtun for his contributions to the first prototype of the front end, as well as Arthur Brack and Anett Hoppe for their contributions to NLP tasks.
\end{acks}

\bibliographystyle{ACM-Reference-Format}
\bibliography{bibliography}

\end{document}

%% file: body.tex
\section{Introduction}
Documents are central to scholarly communication. In fact, nowadays almost all research findings are communicated by means of digital scholarly articles. However, it is difficult to automatically process scholarly knowledge communicated in this form. The content of articles can be indexed and exploited for search and mining, but knowledge represented in form of text, images, diagrams, tables, or mathematical formulas cannot be easily processed and analysed automatically. The primary machine-supported tasks are largely limited to classic information retrieval. Current scholarly knowledge curation, publishing and processing does not exploit modern information systems and technologies to their full potential \cite{journals/isf/Hars01}. As a consequence, the global scholarly knowledge base continues to be little more than a distributed digital document repository.

The key issue is that digital scholarly articles are mere analogues of their print relatives. In the words of Van de Sompel and Lagoze \cite{sompel09aboard} dating back to 2009 and earlier: ``The current scholarly communication system is nothing but a scanned copy of the paper-based system.'' A further decade has gone by and this observation continues to be true. 

Print and digital media suffer from similar challenges. However, given the dramatic increase in output volume and travel velocity of modern research \cite{bornmann15growth} as well as the advancements in information technologies achieved over the past decades, the challenges are more obvious and urgent today than at any time in the past. 

Scholarly knowledge remains as ambiguous and difficult to reproduce~\cite{bosman_commons2017} in digital as it used to be in print. Moreover, addressing modern societal problems relies on interdisciplinary research. Answers to such problems are debated in scholarly discourse spanning often dozens and sometimes hundreds of articles \cite{AlroeSecondOrderScience}. While citation does link articles, their contents are hardly interlinked and generally not machine actionable. Therefore, processing scholarly knowledge remains a manual, and tedious task.

Furthermore, document-based scholarly communication stands in stark contrast to the digital transformation seen in recent years in other information rich publishing and communication services. Examples include encyclopedia, mail order catalogs, street maps or phone books. For these services, traditional document-based publication was not just digitized but has seen the development of completely new means of information organization and access. The striking difference with scholarly communication is that these digital services are not mere digital versions of their print analogues but entirely novel approaches to information organization, access, sharing, collaborative generation and processing.

There is an urgent need for a more flexible, fine-grained, context sensitive and machine actionable representation of scholarly knowledge and corresponding infrastructure for knowledge curation, publishing and processing~\cite{mcaa_policy_2019,publications7020034}. We suggest that representing scholarly knowledge as structured, interlinked, and semantically rich knowledge graphs is a key element of a technical infrastructure~\cite{auer18towards}. While the technology for representing, managing and processing scholarly knowledge in such form is largely in place, we argue that one of the most pressing concerns is how scholarly knowledge can be acquired as it is generated along the research lifecycle, primarily during the \emph{Conducting} step \cite{Vaughan_2013}.

In this article, we introduce the Open Research Knowledge Graph (ORKG) as an infrastructure for the acquisition, curation, publication and processing of semantic scholarly knowledge. We present, evaluate and discuss ORKG based scholarly knowledge acquisition using crowdsourcing and text mining techniques as well as knowledge curation, publication and processing. The alpha release of the ORKG is available online\footnote{\url{https://labs.tib.eu/orkg/}}. Users can provide feedback on issues and features, guide future development with requirements, contribute to implementation and last but not least populate the ORKG with content.

\section{Problem Statement}
\label{sec:Problem}
We illustrate the problem with an example from life sciences. When searching for publications on the popular Genome editing method CRISPR/Cas in scholarly search engines we obtain a vast amount of search results. Google Scholar, for example, currently returns more than $50\,000$ results, when searching for the search string `CRISPR Cas'. Furthermore, research questions often require complex queries relating numerous concepts. Examples for CRISPR/Cas include: How good is CRISPR/Cas w.r.t. precision, safety, cost? How is genome editing applied to insects? Who has applied CRISPR/Cas to butterflies? Even if we include the word `butterfly' to the search query we still obtain more than 600 results, many of which are not relevant. Furthermore, relevant results might not be included (e.g., due to the fact that the technical term for butterfly is Lepidoptera, which combined with `CRISPR Cas' returns over $1\,000$ results). Finally, virtually nothing about the returned scholarly knowledge in the returned documents is machine actionable: human experts are required to further process the results.

We argue that keyword-based information retrieval and document-based results no longer adequately meet the requirements of research communities in modern scholarly knowledge infrastructures \cite{edwards12knowledge} and processing of scholarly knowledge in the digital age. Furthermore, we suggest that automated techniques to identify concepts, relations and instances in text \cite{Singh:2018:WRW:3178876.3186023}, despite decades of research, do not and unlikely will reach a sufficiently high granularity and accuracy for useful applications.
Automated techniques applied on published legacy documents need to be complemented with techniques that acquire scholarly knowledge in machine actionable form as knowledge is generated along the research lifecycle. As Mons suggested \cite{mons05gene}, we may fundamentally question ``Why bury [information in plain text] first and then mine it again.''

This article tackles the following research questions:
\begin{itemize}
    \item Are authors willing to contribute structured descriptions of the key research contribution(s) published in their articles using a fit-for-purpose infrastructure, and what is the user acceptance of the infrastructure?
    \item Can the infrastructure effectively integrate crowdsourcing and automated techniques for multi-modal scholarly knowledge acquisition?
\end{itemize}

\section{Related Work}
\label{sec:RW}

Representing encyclopedic and factual knowledge in machine actionable form is increasingly feasible. This is underscored by knowledge graphs such as Wikidata~\cite{vrandecic14wikidata}, domain-specific knowledge graphs \cite{cheatham18geolink} as well as industrial initiatives at Google, IBM, Bing, BBC, Thomson Reuters, Springer Nature, among others.

In the context of scholarly communication and its operational infrastructure the focus has so far been on representing, managing and linking \emph{metadata} about articles, people, data and other relevant entities. The Research Graph \cite{aryani17researchgraph} is a prominent example of an effort that aims to link publications, datasets and researchers. The Scholix project \cite{burton17scholix}, driven by a corresponding Research Data Alliance working group and associated organizations, standardized the information about the links between scholarly literature and data exchanged among publishers, data repositories, and infrastructures such as DataCite, Crossref, OpenAIRE and PANGAEA. The Research Objects~\cite{bechhofer13researchobjects} project proposed a machine readable abstract structure that relates the products of a research investigation, including articles, data and other research artifacts. The RMap Project \cite{hanson15rmap} aims at preserving ``the many-to-many complex relationships among scholarly publications and their underlying data.'' Sadeghi et al.~\cite{sadeghi17integration} proposed to integrate bibliographic metadata in a knowledge graph.

Some initiatives such as the Semantic Publishing and Referencing (SPAR) Ontologies \cite{peroni14spar} and the Journal Article Tag Suite~\cite{dsm15} extended the representation to \emph{document structure} and more fine-grained elements. Others proposed comprehensive \emph{conceptual models} for scholarly knowledge that capture problems, methods, theories, statements, concepts and their relations \cite{journals/isf/Hars01,de2006modeling,brodaric08skiing,meister17towards}. Allen, R.B. \cite{journals/corr/abs-1209-0036,journals/corr/abs-1708-08423} explored issues related to implementing entire research reports as structured knowledge bases. Fathalla et al. \cite{fathalla17towards} proposed to semantically represent key findings of survey articles by describing research problems, approaches, implementations and evaluations. Nanopublications~\cite{groth2010anatomy} is a further approach to describe scholarly knowledge in structured form. Natural Language Processing based Semantic Scholar \cite{ammar18semanticscholar} and the Machine Learning focused \url{paperswithcode.com} (PWC) are related systems. Key distinguishing aspects among these systems and the ORKG are the granularity of acquired knowledge (specifically, article bibliographic metadata vs. the materials and methods used and results obtained) and the methods used to acquire knowledge (specifically, automated techniques vs. crowdsourcing).

There has been some work on enriching documents with semantic annotations. Examples include Dokie.li~\cite{capadisli17dokieli}, RASH~\cite{peroni17rash} or MicroPublications~\cite{clark14micro} for HTML and SALT~\cite{groza07salt} for \LaTeX.  Other efforts focused on developing ontologies for representing scholarly knowledge in specific domains, e.g. mathematics~\cite{lange13ontologies}.

A knowledge graph for research as proposed here must build, integrate and further advance these and other related efforts and, most importantly, translate what has been proposed so far in isolated prototypes into operational and sustained scholarly infrastructure. There has been work on some pieces but the puzzle has obviously not been solved or more of scientific knowledge would be available today in structured form.

\section{Open Research Knowledge Graph}
\label{sec:ORKG}

We propose to leverage knowledge graphs to represent scholarly knowledge communicated in the literature. We call this knowledge graph the Open Research Knowledge Graph\footnote{\url{http://orkg.org}} (ORKG). Crucially, the proposed knowledge graph does not merely contain (bibliographic) metadata (e.g., about articles, authors, institutions) but semantic (i.e., machine actionable) descriptions of scholarly knowledge.

\begin{figure}[tb]
    \centering
    \includegraphics[width=\columnwidth]{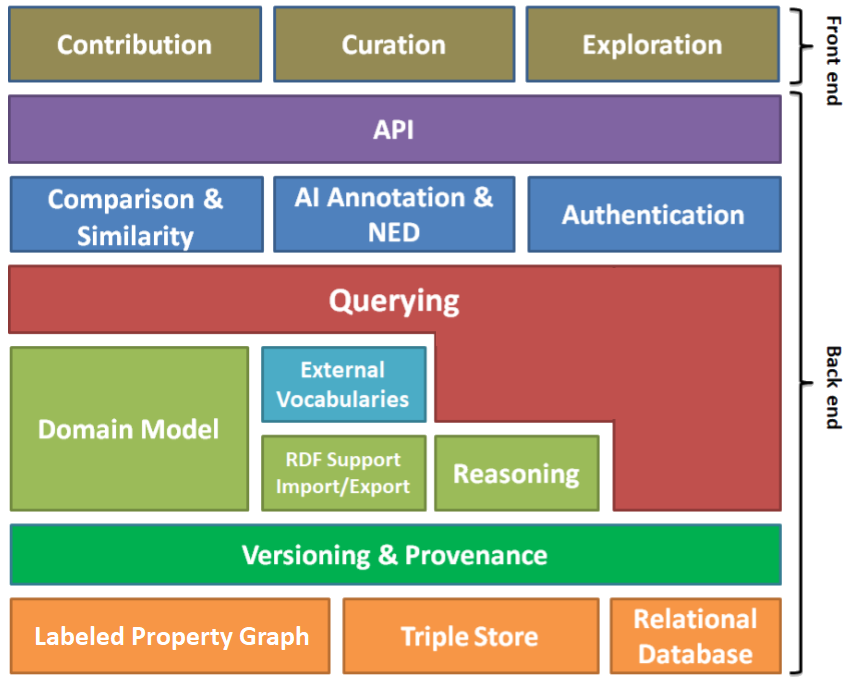}
    \caption{ORKG layered architecture from data persistence to services.}
    \label{fig:architecture}
\end{figure}

\subsection{Architecture}
\label{ssec:architecture}
The infrastructure design follows a classical layered architecture. As depicted in Figure \ref{fig:architecture}, a persistence layer abstracts data storage implemented by labeled property graph (LPG), triple store, and relational database storage technology, each serving specific purposes. Versioning and provenance handles tracking changes to stored data.

The domain model specifies \texttt{ResearchContribution}, the core information object of the ORKG. A \texttt{ResearchContribution} relates the \texttt{ResearchProblem} addressed by the contribution with the \texttt{ResearchMethod} and (at least one) \texttt{ResearchResult}. Currently, we do not further constrain the description of these resources. Users can adopt arbitrary third-party vocabularies to describe problems, methods, and results. For instance, users could  use the Ontology for Biomedical Investigations as a vocabulary to describe statistical hypothesis tests.

Research contributions are represented by means of a graph data model. Similarly to the Research Description Framework~\cite{brickley2014rdfs} (RDF), the data model is thus centered around the concept of a statement, a triple consisting of two nodes (resources) connected by a directed edge. In contrast to RDF, the data model allows to uniquely identify instances of edges and to qualify these instances (i.e., annotate edges and statements). As metadata of statements, provenance information is a concrete and relevant application of such annotation.

RDF import and export enables data synchronization between LPG and triple store, which enables SPARQL and reasoning. Querying handles the requests by services for reading, updating, and creating content in databases. The following layer is for modules that implement infrastructure features such as authentication or comparison and similarity computation. The REST API acts as the connector between features and services for scholarly knowledge contribution, curation and exploration.

ORKG users in author, researcher, reviewer or curator roles interact differently with its services. Exploration services such as State-of-the-Art comparisons are useful in particular for researchers and reviewers. Contribution services are primarily for authors who intend to contribute content. Curation services are designed for domain specialists more broadly to include for instance subject librarians who support quality control, enrichment and other content organization activities.

\subsection{Features}
\label{ssec:features}

The ORKG services are underpinned by numerous features that, individually or in combination, enable services. We present the most important current features next.

\paragraph{\textbf{State-of-the-Art (SOTA) comparison}.} SOTA comparison extracts similar information shared by user selected research contributions and presents comparisons in tabular form. Such comparisons rely on extracting the set of semantically similar predicates among compared contributions. 

We use  FastText~\cite{bojanowski2017enriching} word embeddings to generate a similarity matrix $\gamma$

\begin{equation}
    \gamma = \left [ cos(\overrightarrow{p_i}, \overrightarrow{p_j}) \right ]
    \label{eq:similarity_matrix}
\end{equation}

with the cosine similarity of vector embeddings for predicate pairs $(p_i, p_j) \in \mathcal{R}$, whereby $\mathcal{R}$ is the set of all research contributions.

Furthermore, we create a mask matrix $\Phi$ that selects predicates of contributions $c_i \in \mathcal {C}$, whereby $\mathcal{C}$ is the set of research contributions to be compared. Formally,

\begin{equation}
    \Phi_{i,j} = \begin{cases}
1 & \text{ if } p_{j} \in c_i \\ 
0 & \text{ otherwise }  
\end{cases}
    \label{eq:mask_matrix}
\end{equation}

Next, for each selected predicate $p$ we create the matrix $\varphi$ that slices $\Phi$ to include only similar predicates. Formally,

\begin{equation}
\varphi_{i,j} =(\Phi_{i,j})_{\substack{c_i \in \mathcal{C}\\ p_j \in sim(p) }}
\label{eq:slice_mask}
\end{equation}

where $sim(p)$ is the set of predicates with similarity values $\gamma[p] \geq T = 0.9$ with predicate $p$. The threshold $T$ is computed empirically. Finally, $\varphi$ is used to efficiently compute the common set of predicates and their frequency. 

\paragraph{\textbf{Contribution similarity}.} Contribution similarity is a feature used to explore related work, find or recommend comparable research contributions. The sub-graphs $G(r_i)$ for each research contribution $r_i \in \mathcal{R}$ are converted into document $D$ by concatenating the labels of subject $s$, predicate $p$, and object $o$, of all statements $(s,p,o) \in G(r_i)$. We then use $TF/iDF$~\cite{ramos2003using} to index and retrieve the most similar contributions with respect to some query $q$. Queries are constructed in the same manner as documents $D$.

\paragraph{\textbf{Automated Information Extraction}.} The ORKG uses machine learning for automated extraction of scientific knowledge from literature. Of particular interest are the NLP tasks named entity recognition as well as named entity classification and linking.

As a first step, we trained a neural network based machine learning model for named entity recognition using in-house developed annotations on the Elsevier Labs corpus of Science, Technology, and Medicine\footnote{https://github.com/elsevierlabs/OA-STM-Corpus} (STM) for the following generic concepts: process, method, material and data. 
We use the Beltagy et al.~\cite{Beltagy2019SciBERTPC} Named Entity Recognition task-specific neural architecture atop pretrained SciBERT embeddings with a CRF-based sequence tag decoder~\cite{Ma2016EndtoendSL}.

Linking scientific knowledge to existing knowledge graphs including those from the open domain such as DBpedia~\cite{auer2007dbpedia} as well as domain specific graphs such as ULMS~\cite{bodenreider2004unified} is another important feature. Most importantly, such linking enables semi-automated enrichment of research contributions.

\subsection{Implementation}
\label{ssec:implementation}

The infrastructure consists of two main components: the \emph{back end} with the business logic, system features and a set of APIs used by services and third party apps; and the \emph{front end}, i.e. the User Interface (UI).

\paragraph{\textbf{Back end}.} The back end is written in Kotlin~\cite{Kotlin}, within the Spring Boot 2 framework. The data is stored in a Neo4j Labeled Property Graph (LPG) database accessed via the Spring Data Neo4j's Object Graph Mapper (OGM). Data can be queried using Cypher, Neo4j's native query language. The back end is exposed via a JSON RESTful API accessed by applications, including the ORKG front end. A technical documentation of the current API specification is available online\footnote{\url{http://tib.eu/c28v}}.

Data can be exported to RDF via the Neo4j Semantics extension\footnote{\url{https://github.com/jbarrasa/neosemantics}}. Due to the differences between our graph model and RDF, a ``semantification'' needs to occur. Most importantly, the ORKG back end auto-generates URIs. Mapping (or changing) these URIs to an existing ontology must be done manually. The Semantics extension also allows importing RDF data and ontologies. The Web Ontology Language is, however, not fully supported.

In order to enrich ORKG data, we support linking to data from other sources. Of particular interest are systems that collect, curate and publish bibliographic metadata or data about entities relevant to scholarly communication, such as people and organizations. Rather than collecting such metadata ourselves we thus link and integrate with relevant systems (e.g., DataCite, Crossref) and their data via identifiers such as DOI, ORCID or ISNI.

\begin{figure}[tb]
    \centering
    \includegraphics[width=\columnwidth]{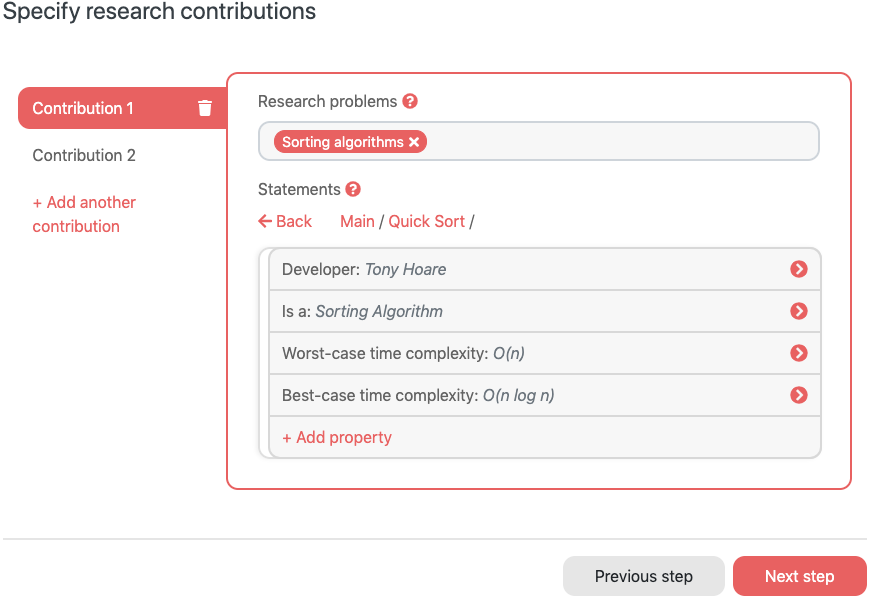}
    \caption{Snapshot of the create research contribution wizard in the front end.}
    \label{fig:ui-step-3}
\end{figure}

\begin{figure}[tb]
    \centering
	\includegraphics[width=\columnwidth]{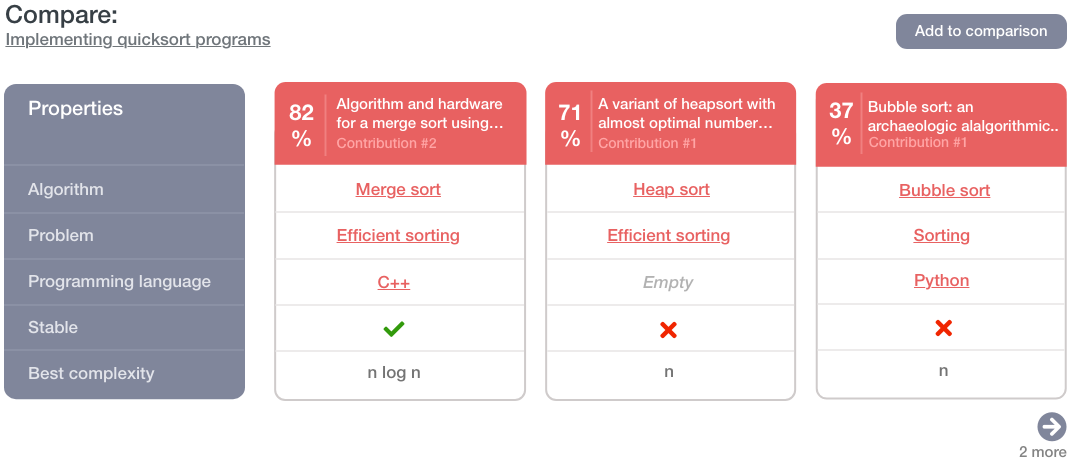}
	\caption{State-of-the-Art comparison between ``Quick Sort'' and other similar research contributions.} \label{fig:ui-state-of-the-art}
\end{figure}

\paragraph{\textbf{Front end}.} The ORKG front end is a Web-based tool for, among other users, researchers and librarians and supports searching, exploring, creating and modifying research contributions. Figure \ref{fig:ui-step-3} depicts the wizard that guides users in creating research contributions. Figure \ref{fig:ui-state-of-the-art} depicts comparing \textit{Quick Sort} to other research contributions. This service combines the similarity and comparison features to deliver the depicted table. The table can be shared via a persistent link and exported to different formats, including \LaTeX, CSV and PDF.

The front end was built with the following two key requirements in mind: (1) \textit{Usability} to enable a broad range of users, in particular researchers across disciplines, to contribute, curate and explore research contributions; and (2) \textit{Flexibility} to enable maximum degree of freedom in describing research contributions.

It is implemented according to the ES6 standard of JavaScript using the React\footnote{\url{https://reactjs.org/}} framework. The Bootstrap\footnote{\url{https://getbootstrap.com}} framework is used for responsive interface components. 

The front end design takes great care to deliver the best possible overall experience for a broad range of users, in particular researchers not familiar with knowledge graph technology. User evaluations are a key instrument to continually validate the development and understand requirements.

\begin{savenotes}
\begin{table*}[ht]
\centering
\caption{Overview of answers to the key aspects covered by the evaluation questionnaire and other metrics recorded during the interviews.}
\begin{adjustbox}{scale=.8}
\begin{tabular}{cccccccccc}
\hline
\multirow{2}{*}{\textbf{\begin{tabular}[c]{@{}c@{}}Participant\\ Nr\end{tabular}}} & \textbf{Navigation}                                          & \textbf{Terminology}                                             & \textbf{Auto Complete}                                     & \textbf{Guidance Needed}                                   & \textbf{Suggest To Others}                                & \textbf{UI likeness}                                    & \multicolumn{1}{l}{\textbf{Time}} & \multirow{2}{*}{\textbf{\begin{tabular}[c]{@{}c@{}}Number of\\ Triples\end{tabular}}} & \multirow{2}{*}{\textbf{\begin{tabular}[c]{@{}c@{}}Properly\\ Structured\footnote{Meaning that the participant added information following some organization scheme, including distinct and possibly further refined problem, method and result.}\end{tabular}}} \\ \cline{2-8}
                                                                                   & \begin{tabular}[c]{@{}c@{}}5 = Very\\ intuitive\end{tabular} & \begin{tabular}[c]{@{}c@{}}5 = Easy to\\ understand\end{tabular} & \begin{tabular}[c]{@{}c@{}}5 = Very\\ helpful\end{tabular} & \begin{tabular}[c]{@{}c@{}}5 = All\\ the time\end{tabular} & \begin{tabular}[c]{@{}c@{}}9 = Very\\ likely\end{tabular} & \begin{tabular}[c]{@{}c@{}}9 = Very\\ much\end{tabular} & in mins                           &                                                                                       &                                                                                         \\ \hline
1                                                                                  & 4                                                            & 4                                                                & 5                                                          & 3                                                          & 2                                                         & 66                                                      & 16                                & 56                                                                                    & \cmark                                                                                       \\
2                                                                                  & 2                                                            & 3                                                                & 5                                                          & 4                                                          & 8                                                         & 7                                                       & 19                                & 35                                                                                    & \xmark                                                                                       \\
3                                                                                  & 4                                                            & 5                                                                & 5                                                          & 3                                                          & 9                                                         & 7                                                       & 15                                & 81                                                                                    & \cmark                                                                                       \\
4                                                                                  & 3                                                            & 3                                                                & 5                                                          & 3                                                          & 6                                                         & 7                                                       & 13                                & 27                                                                                    & \xmark                                                                                       \\
5                                                                                  & 4                                                            & 3                                                                & 5                                                          & 3                                                          & 6                                                         & 8                                                       & 14                                & 48                                                                                    & \xmark                                                                                       \\
6                                                                                  & 4                                                            & 3                                                                & 5                                                          & 3                                                          & 8                                                         & 9                                                       & 13                                & 23                                                                                    & \xmark                                                                                       \\
7                                                                                  & 3                                                            & 4                                                                & 5                                                          & 3                                                          & 7                                                         & 6                                                       & 19                                & 57                                                                                    & \cmark                                                                                       \\
8                                                                                  & 3                                                            & 2                                                                & 4                                                          & 3                                                          & 8                                                         & 6                                                       & 13                                & 56                                                                                    & \cmark                                                                                       \\
9                                                                                  & 4                                                            & 5                                                                & 3                                                          & 3                                                          & 7                                                         & 5                                                       & 14                                & 55                                                                                    & \xmark                                                                                       \\
10                                                                                 & 4                                                            & 5                                                                & 5                                                          & 1                                                          & 8                                                         & 8                                                       & 22                                & 68                                                                                    & \cmark                                                                                       \\
11                                                                                 & 4                                                            & 5                                                                & 5                                                          & 1                                                          & 8                                                         & 8                                                       & 20                                & 55                                                                                    & \cmark                                                                                       \\
12                                                                                 & -                                                            & -                                                                & -                                                          & -                                                          & -                                                         & -                                                       & 21                                & 71                                                                                    & \xmark                                                                                       \\ \hline
\textbf{Average}                                                                   & \textbf{4}                                                   & \textbf{4}                                                       & \textbf{5}                                                 & \textbf{3}                                                 & \textbf{7}                                                & \textbf{7}                                              & \textbf{17}                       & \textbf{53}                                                                           & \textbf{-}                                                                              \\ \hline
\end{tabular}
\end{adjustbox}
\label{tab:survey-results}
\end{table*}
\end{savenotes}

\section{Evaluations}
\label{sec:Evaluation}

The ORKG infrastructure, its services, features, performance and usability are continually evaluated to inform the next iteration and future developments. Among other preliminary evaluations and results, we present here the \emph{first} front end user evaluation, which informed the second iteration of front end development, presented in this paper. 

\subsection{Front end Evaluation}
\label{ssec:frontend-eval}
Following a qualitative approach, the evaluation of the first iteration of front end development aimed to determine user performance, identify major (positive and negative) aspects, and user acceptance/perception of the system. The evaluation process had two components: (1) instructed interaction sessions and (2) a short evaluation questionnaire. This evaluation resulted in data relevant to our first research question.

We conducted instructed interaction sessions with 12 authors of articles presented at the DILS2018\footnote{\url{https://www.springer.com/us/book/9783030060152}} conference. The aim of these sessions was to get first-hand observations and feedback. The sessions were conducted with the support of two instructors. At the start of each session, the instructor briefly explained the underlying principles of the infrastructure, including how it works and what is required from authors to complete the task, i.e. create a structured description of the key research contribution in their article presented at the conference. Then, participants engaged with the system without further guidance from the instructor. However, at any time they could ask the instructor for assistance. For each participant, we recorded the time required to complete the task (to determine the mean duration of a session), the instructor's notes and the participant's comments.

In addition to the instructed interaction sessions, participants were invited to complete a short evaluation questionnaire. The questionnaire is available online\footnote{\url{https://doi.org/10.5281/zenodo.2549918}}. Its aim was to collect further insights into user experience. Since the quantity of collected data was insufficient to establish any correlational or causal relationship \cite{jansen_logic_2010}, the questionnaire was treated as a qualitative instrument. The paper-based questionnaire consisted of 11 questions. These were designed to capture participant thoughts regarding the positive and negative aspects of the system following their instructed interaction session. Participants completed their questionnaire after the instructed interaction session. All 12 participants answered the questionnaire. The interaction notes, participant comments and the time recordings were collected together with questionnaire responses and analysed in light of our research questions. 

A dataset summarizing the research contributions collected in the experiment is available online\footnote{\url{https://doi.org/10.5281/zenodo.3340954}}. The data is grouped into four main categories. \textit{Research Problem} describes the main question or issue addressed by the research contribution. Participants used a variety of properties to describe the problem, e.g. problem, addresses, subject, proposes and topic. \textit{Approach} describes the solution taken by the authors. Properties used included approach, uses, prospective work, method, focus and algorithm. \textit{Implementation} \& \textit{Evaluation} were the most comprehensively described aspects, arguably because it was easier for participants to describe technical details compared to describing the problem or the approach.

In summary, 75\% of the participants found the front end developed in the first iteration fairly intuitive and easy to use. Among the participants, 80\% needed guidance only at the beginning while 10\% did not need guidance. The time required to complete the task was 17 minutes on average, with a minimum of 13 minutes and a maximum of 22 minutes. Five out of twelve participants suggested to make the front end more keyboard-friendly to ease the creation of research contributions. As participant \#3 stated: ``More description in advance can be helpful.'' Two participants commented that the navigation process throughout the system is complicated for first-time users and suggested alternative approaches. As an example, participant \#5 suggested to ``Use breadcrumbs to navigate.''

Four participants wished for a visualisation (i.e., graph chart) to be available when creating new research contributions. For instance, participant \#1 commented that ``It could be helpful to show a local view of the graph while editing.'' This type of visualisation could facilitate comprehension and ease curation. Another participant suggested to integrate a document (PDF) viewer and highlight relevant passages or phrases. Participant \#4 noted that ``If I could highlight the passages directly in the paper and add predicates there, it would be more intuitive and save time.'' 

Further details of the questionnaire, including participant ratings on main issues, are summarized in Table \ref{tab:survey-results}. While the cohort of participants was too small for statistically significant conclusions, these results provided a number of important suggestions that informed the second iteration of front end development, which is presented in this paper and will be evaluated in future conferences.  

\begin{table}[t]
\caption{Time (in seconds) needed to perform State-of-the-Art comparisons with 2-8 research contributions using the baseline and ORKG approaches.}
\begin{adjustbox}{scale=.8}
\begin{tabular}{l|lllllll}
& \multicolumn{7}{c}{\textbf{Number of compared research contributions}} \\
\hline
                  & \textbf{2} & \textbf{3} & \textbf{4} & \textbf{5} & \textbf{6} & \textbf{7} & \textbf{8} \\ \hline
\textbf{Baseline} & 0.00026         & 0.1714          & 0.763          & 4.99            & 112.74          & 1772.8          & 14421           \\
\textbf{ORKG}     & 0.0035          & 0.0013          & 0.01158         & 0.02            & 0.0206          & 0.0189          & 0.0204         
\end{tabular}
\end{adjustbox}
\label{tab:sota-eval}
\end{table}

\subsection{Other Evaluations}
We have performed preliminary evaluations also of other components of the ORKG infrastructure. The experimental setup for these evaluations is an Ubuntu 18.04 machine with Intel Xeon CPUs $12 \times 3.60$ GHz and 64 GB memory.

\paragraph{\textbf{Comparison feature}.} With respect to the State-of-the-Art comparison feature, we compared our approach in ORKG with the baseline approach, which uses brute force to find the most similar predicates and thus checks every possible predicate combination. Table \ref{tab:sota-eval} shows the time needed to perform the comparison for the baseline approach and for the approach we implemented and presented above. As the results suggest, our approach clearly outperforms the baseline and the performance gain can be attributed to more efficient retrieval. The experiment is limited to 8 contributions because the baseline approach does not scale to larger sets.

\paragraph{\textbf{Scalability}.} For horizontal scalability, the infrastructure containerizes applications. We also tested the vertical scalability in terms of response time. For this, we created a synthetic dataset of papers. Each paper includes one research contribution described by three statements. The generated dataset contains 10 million papers or 100 million nodes. We tested the system with variable numbers of papers and the average response time to fetch a single paper with its related research contribution is 60 ms. This suggests that the infrastructure can handle large amounts of scholarly knowledge.

\begin{figure}[H]
    \centering
	\includegraphics[width=.85\columnwidth]{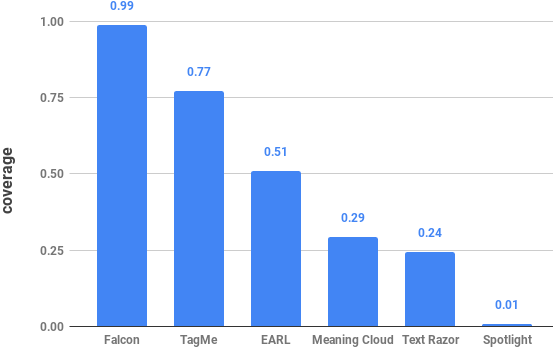}
	\caption{Coverage values of different NED systems over the annotated entities of the STM corpus.}
	\label{fig:coverage}
\end{figure}

\paragraph{\textbf{NED performance}.} We evaluated the performance of a number of existing NED tools on scholarly knowledge, specifically Falcon~\cite{sakor2019old}, DBpedia Spotlight~\cite{journals/tacl/0001RN14}, TagME~\cite{conf/cikm/FerraginaS10}, EARL~\cite{conf/EARL}, TextRazor\footnote{\url{https://www.textrazor.com/docs/rest}} and MeaningCloud\footnote{\url{https://www.meaningcloud.com/developer}}. These tools were used to link to entities from Wikidata and DBpedia. We used the annotated entities from the STM corpus as the experimental data. However, since there is no gold standard for the dataset, we only computed the coverage metric $\zeta = \textit{\# of linked entities}$ divided by $\textit{\# of all entities}$. Figure \ref{fig:coverage} summarizes the coverage percentage for the evaluated tools. The results suggest that Falcon is most promising. 

\section{Discussion and Future Work}
\label{sec:Discussion}

We model scholarly knowledge communicated in the scholarly literature following the abstract concept of \texttt{ResearchContribution} with a simplistic concept description. This is especially true if compared to some conceptual models of scholarly knowledge that can be found in the literature, e.g. the one by Hars~\cite{journals/isf/Hars01}. While comprehensive models are surely appealing to information systems, e.g. for the advanced applications they can enable, we think that populating a database with a complex conceptual model is a significant challenge, one that remains unaddressed. Conscious of this challenge, for the time being we opted for a simplistic model with lower barriers for content creation.

A further and more fundamental concern is the granularity with which scholarly knowledge can realistically be acquired, beyond which the problem becomes intractable. How graph data models and management systems can be employed to represent and manage granular and large amounts of interconnected scholarly knowledge as well as knowledge evolution is another open issue.

With the evaluation of the first iteration of front end development we were able to scrutinize various aspects of the infrastructure and obtain feedback on user interaction, experience and system acceptance. Our results show that the infrastructure meets key requirements: it is easy to use and users can flexibly create and curate research contributions. The results of the questionnaire (Table \ref{tab:survey-results}) show that with the exception of `Guidance Needed', all aspects were evaluated above average (i.e., positively). The results suggest that guidance from an external instructor is not needed, reinforcing the usability requirement of the front end. All case study participants displayed an interest in the ORKG and provided valuable input on what should be changed, added or removed. Furthermore, participants suggested to integrate the infrastructure with (digital) libraries, universities, and other institutions.

Based on these preliminary findings, we suggest that authors are willing to provide structured descriptions of the key contribution published in their articles. The practicability of crowdsourcing such descriptions at scale and in heterogeneous research communities needs further research and development.

\begin{figure}[tb]
    \centering
    \includegraphics[width=\columnwidth]{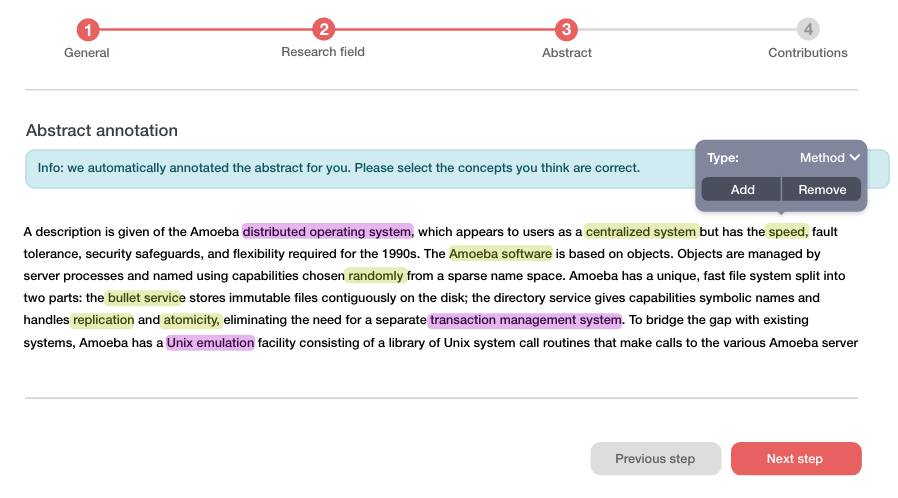}
    \caption{Mock-up of automated information extraction feature integrated into the research contribution workflow of the infrastructure.}
    \label{fig:annotation}
\end{figure}

In support of our second research question, we argue that the presented infrastructure prototypes the integration of both crowdsourced and automated modes of semantic scholarly knowledge acquisition and curation. With the NLP task models and experiments as well as designs for their integration in the front end, we suggest that the building blocks for integrating automated techniques in user interfaces for manual knowledge acquisition and curation are in place (Figure \ref{fig:annotation}). In addition to these two modes, \cite{stocker18ri} have proposed a third mode whereby scholarly knowledge is acquired as it is generated during the research lifecycle, specifically during data analysis by integrating with computational environments such as Jupyter.

Next steps for the ORKG include importing semi-structured scholarly knowledge from other sources, such as PWC. Such existing content is crucial to build a sizable database to use for future development, testing and demonstration. 

User authentication and versioning to track the evolution of curated scholarly knowledge are additional important features on the ORKG development roadmap. Furthermore, we are working on integrating scholarly knowledge (graph) visualisation approaches in the front end to support novel and advanced exploration. Interoperability with related data models such as nanopublications will be addressed. Furthermore, we plan to integrate a discussion feature that will enable debating existing scholarly knowledge, e.g. for knowledge correctness. 

Finally, we will further work on automated information extraction, to support, guide and ease information entering. Specifically, we are developing a front end feature that identifies and displays relevant conceptual zones in article content. When a user enters information about, say, the problem addressed by the research contribution then the interface will present problem-related text extracted from the article. This feature will ease entering information, possibly by enabling users to simply highlight the relevant part in the text.

\section{Conclusion}
\label{sec:Conclusions}

This article described the first steps of a larger research and development agenda that aims to enhance document-based scholarly communication with semantic representations of communicated scholarly knowledge. In other words, we aim to bring scholarly communication to the technical standards of the 21st century and the age of modern digital libraries. We presented the architecture of the proposed infrastructure as well as a first implementation. The front end has seen substantial development, driven by earlier user feedback. We have reported here the results of the user evaluation to underpin the development of the current front end, which was recently released for public view and use. By integrating crowdsourcing and automated techniques in natural language processing, initial steps were also taken and evaluated that advance multi-modal scholarly knowledge acquisition using the ORKG.